\documentclass[runningheads]{llncs}
\usepackage{graphicx}
\usepackage{url}
\begin{document}
\title{Metadata Extraction from Raw Astroparticle Data of TAIGA Experiment}
\author{
Igor Bychkov\inst{1} \and
Julia Dubenskaya\inst{2} \and
Elena Korosteleva\inst{2} \and
Alexandr Kryukov\inst{2} \and \\
Andrey Mikhailov\inst{1} \and 
Minh-Duc Nguyen\inst{2} \and 
Alexey Shigarov\inst{1,3}
}
\authorrunning{I. Bychkov et al.}
\institute{Matrosov Institute for System Dynamics and Control Theory, Siberian Branch of Russian Academy of Sciences, Lermontov St. 134, Irkutsk, Russia\\
\email{shigarov@icc.ru}\\
\and
Skobeltsyn Institute of Nuclear Physics, Lomonosov Moscow State University, Leninskiye Gory 1(2), Moscow, Russia\\
\email{kryukov@theory.sinp.msu.ru}\\
\and
Institute of Mathematics, Economics and Informatics, Irkutsk State University, Gagarin Blvd. 20, Irkutsk, Russia
}
\maketitle              % typeset the header of the contribution
\begin{abstract}

Today, the operating TAIGA (Tunka Advanced Instrument for cosmic rays and Gamma Astronomy) experiment continuously produces and accumulates a large volume of raw astroparticle data.
To be available for the scientific community these data should be well-described and formally characterized.
The use of metadata makes it possible to search for and to aggregate digital objects (e.g. events and runs) by time and equipment through a unified interface to access them.
The important part of the metadata is hidden and scattered in folder/files names and package headers. 
Such metadata should be extracted from binary files, transformed to a unified form of digital objects, and loaded into the catalog.
To address this challenge we developed a concept of the metadata extractor that can be extended by facility-specific extraction modules.
It is designed to automatically collect descriptive metadata from raw data files of all TAIGA formats.

\end{abstract}
\section{Introduction}

Nowadays, large-scale setups used in the experimental astroparticle physics generate a large volume of data.
This trend gives rise to a number of emerging issues of big data management~\cite{Demchenko2012}. 
Some activities should be carried out continuously across all stages of the \textit{astroparticle data life cycle}~\cite{Bychkov2018} 
and \textit{open science}~\cite{David2004,Nosek2015}, the model of free access to data, for the astroparticle physics.

The \textit{Russian-German astroparticle data life cycle initiative}\footnote{\url{https://astroparticle.online}}~\cite{Bychkov2018} aims at developing an open science system 
to support collection, storage, analysis, sharing and reuse of data produced by TAIGA\footnote{\url{https://taiga-experiment.info}} experimental facilities~\cite{Budnev2016}. 
This system is designed to be a common data portal of two independent observatories and at the same time for consolidation of an analysis for astroparticle physics experiments. 

One of the important issues is how to efficiently manage raw astroparticle data to support their availability and reuse in future. 
The long-term preservation of raw data as originally generated is essential for re-running analysis and reproducing research results. 
To be accessible for the scientific community raw data should be well-described by descriptive, structural, and administrative metadata.
The use of such metadata makes it possible to search for and to aggregate raw astroparticle data through a unified interface to access them.

Metadata is useful on all stages of the data life cycle we considered in our initiative~\cite{Bychkov2018}: 
\textit{data availability} covers user requests to data through metadata;
\textit{data analysis} can enrich metadata;
\textit{simulations} can generate metadata;
\textit{open access} is implemented by metadata describing ownership and rights;
\textit{education in data science} uses metadata of educational resources;
\textit{data archiving} provides a long-term data preservation alongside metadata.

Currently, the TAIGA experimental facilities use five unique binary file formats for representing raw data 
produced by the gamma ray setups: TAIGA-HiSCORE~\cite{Prosin2016} and TAIGA-IACT~\cite{Kuzmichev2018}, 
and the cosmic ray setups: Tunka-133~\cite{Prosin2016}, Tunka-Grande~\cite{Monkhoev2017}, and Tunka-Rex~\cite{Bezyazeekov2015}. 
They are not accompanied by well-organized metadata. 
Some scattered metadata are hidden in names and package headers of raw data files.
There is neither conventional terminology used in the experiments nor the unified interface for access to the hidden metadata.
The main challenge being considered in this paper is how to extract metadata from raw data.

We propose a concept of the metadata extractor designed to be used in the astroparticle data storage~\cite{Kryukov2018}.
The extractor is aimed to automatically collect descriptive metadata from raw data files of all TAIGA formats and put them into a catalog of the storage.
Its architecture is extensible by facility-specific extraction modules (add-ons)
that can be implemented with a framework for binary data format description such as 
``Kaitai Struct''\footnote{\url{http://kaitai.io}} or ``FlexT''\footnote{\url{http://hmelnov.icc.ru/flext}}.
Extracted metadata should provide searching for and aggregating meaningful chunks of raw data by both time and equipment.

\section{Background}

We define the \textit{metadata extraction} as a characterization of \textit{digital objects} that are concealed by raw data.
The digital objects significant for purposes of our initiative are \textit{events} being registered by detectors.
An event as a digital object is composed of a structured sequence of bits/bytes in binary files.
The sequence of bits/bytes defining such an event can be accessed using a set of unique identifiers represented in package headers.

Our digital object is also characterized by time and equipment of a \textit{run} where it is registered.
These properties are scattered over the folder structure, file names, and package headers.
There are some software tools that can be used for the metadata extraction from binary files.
We consider here two kinds of them: (i) tools for harvesting metadata from binary files and (ii) frameworks for the binary data format description.

%\subsection{Harvesting metadata from binary files}

There are several contemporary tools for harvesting metadata from binary files, including the followings: 
``NLNZ Metadata Extraction Tool''\footnote{\url{http://meta-extractor.sourceforge.net}},  
``JHOVE2''\footnote{\url{https://bitbucket.org/jhove2/main/wiki/Home}},
``FITS''\footnote{\url{https://projects.iq.harvard.edu/fits}}, 
and ``GNU Libextractor''\footnote{\url{https://www.gnu.org/software/libextractor}}.
Typically, such tools support some wide-spread file formats (e.g. JPEG, MP3, ZIP) as a source.
They store extracted metadata in XML, JSON, or delimited text files.
Their functionality can be extended by plug-ins or modules for processing specific binary formats.

A workflow for the characterization of digital objects can include the following steps:
\textit{identification}, i.e. determining a presumptive format used for representation of a digital object;
\textit{validation}, i.e. determining the conformance of a digital object to the identified format;
\textit{extraction}, i.e. deriving metadata of a digital object significant for purposes of classification, analysis, and use;
and \textit{assessment}, i.e. determining the acceptability of a digital object for a specific use.
The architecture of our metadata extractor is designed on the basis of this workflow.

%\subsection{Binary data format description}

The state-of-the-art frameworks, such as ``Kaitai Struct'' or ``FlexT'', provide formal languages for describing binary data formats~\cite{Bychkov2018b}.
They are a satisfactory solution for the issues of raw data documenting, parsing and verifying. 
Our previous work~\cite{Bychkov2018b} demonstrates applicability of binary file format description languages to specify, parse and verify raw data of TAIGA experiment. 
The formal specifications implemented for the five formats of the experiments make it possible 
to automatically generate the source code of libraries for accessing data in one of the target programming languages (e.g. C++, Java, or Python).
They demonstrate good performance and allow us to locate files with corrupted data.

Such frameworks can facilitate the extraction of metadata from binary files.
We use format specifications in the formal languages to implement facility-specific modules that extend the capabilities of our metadata extractor.
Each module relies on a format-oriented data reading library generated automatically from the corresponding specification by the framework.
This allows us to identify the file format, to validate raw data, and to extract the descriptive attributes of digital objects.

Among the existing solutions for describing the binary data format, ``Kaitai Struct'' and ``FlexT'' are the most suitable ones in our case~\cite{Bychkov2018b}.
Both provide the declarative languages for representing file format specifications. 
Similarly, they consider a specification as a set of data type definitions. 
They support bit-oriented data (bit fields) and variant blocks. 
Both allow one to generate the source code of the reading libraries for the raw data formats from specifications.
``FlexT'' language~\cite{Hmelnov2016} is more expressive, but ``Kaitai Struct'' is based on well-known format, namely YAML. 
Moreover, ``Kaitai Struct'' supports more programming languages for the source code generation. 

\section{Metadata of TAIGA raw data}

%\subsection{Hidden metadata}

\begin{figure}[t]
	\centering
		\includegraphics[width=1.0\textwidth]{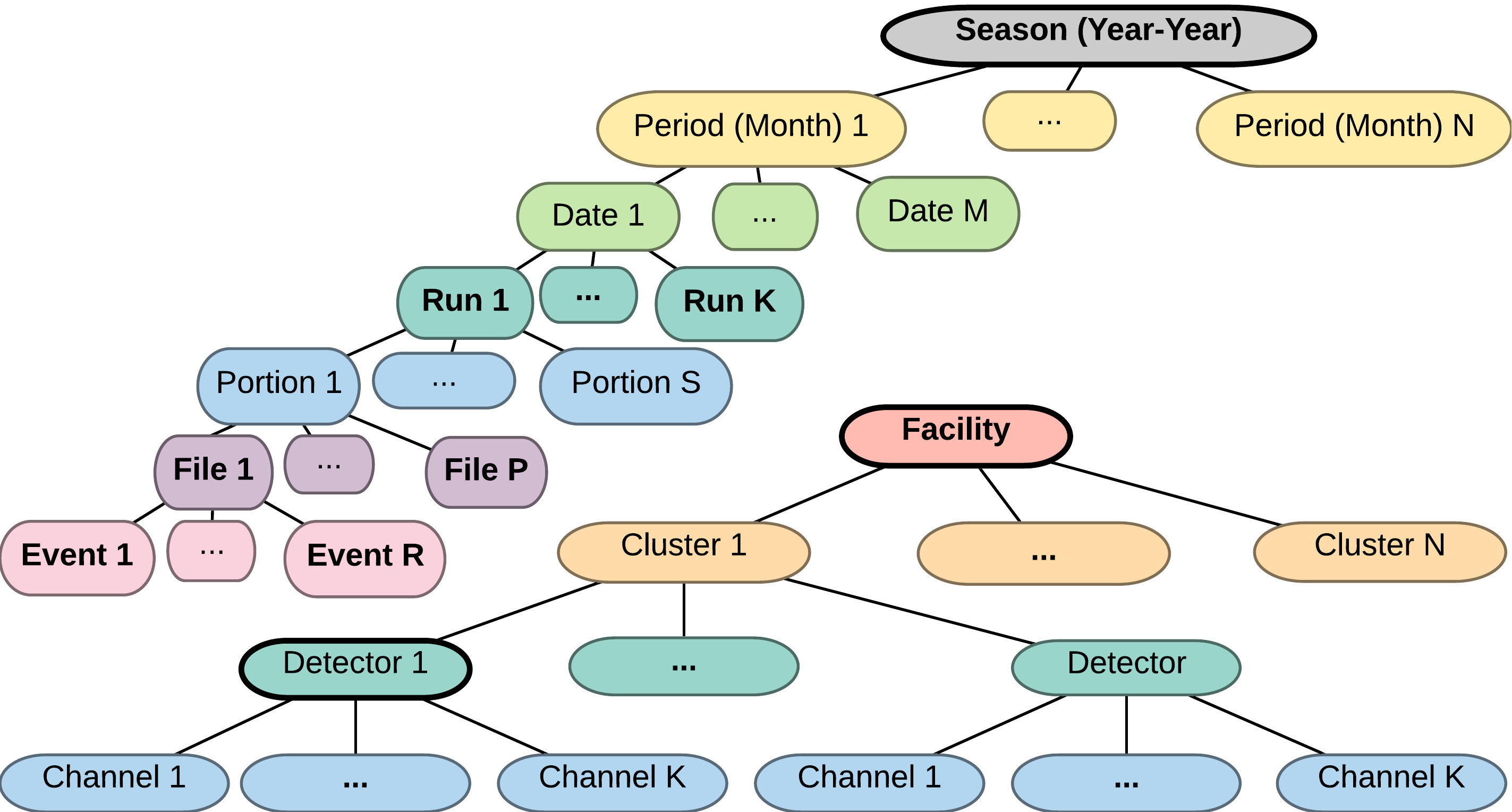}
	\label{fig:md1}
	\caption{Aspects of time and equipment in metadata hidden in TAIGA raw data.}
\end{figure}

%The TAIGA raw data contains 
An important part of the available metadata characterizing events is hidden in the folder/file names and package headers of raw data files.
There are two important dimensions, time and equipment, describing each event registered in a run of a facility (see Fig.~\ref{fig:md1}).
These dimensions define a hierarchical structure of folders with raw data for each facility as follows: 
a season of measuring (the folder ``YYYY-YY'' --- a start and end year), 
a moonless period of measuring in a month (the folder ``mmmYY'' --- a month abbreviation and year),
a night (the folder ``ddmmyy.NN'' --- a date and a number of run in the night), 
a cluster or station (the folder ``fffNNN'' --- a facility and a number),
a raw data file (the folder ``ddmmy.NNN'' a date and a portion of the raw data writing).
Both dimensions are also presented in package headers inside raw data files.
A package header can include a timestamp with an accuracy from millisecond to nanosecond depending on the facility, as well as a detector and channel identifier of the equipment.

\begin{figure}[t]
	\centering
		\includegraphics[width=1.0\textwidth]{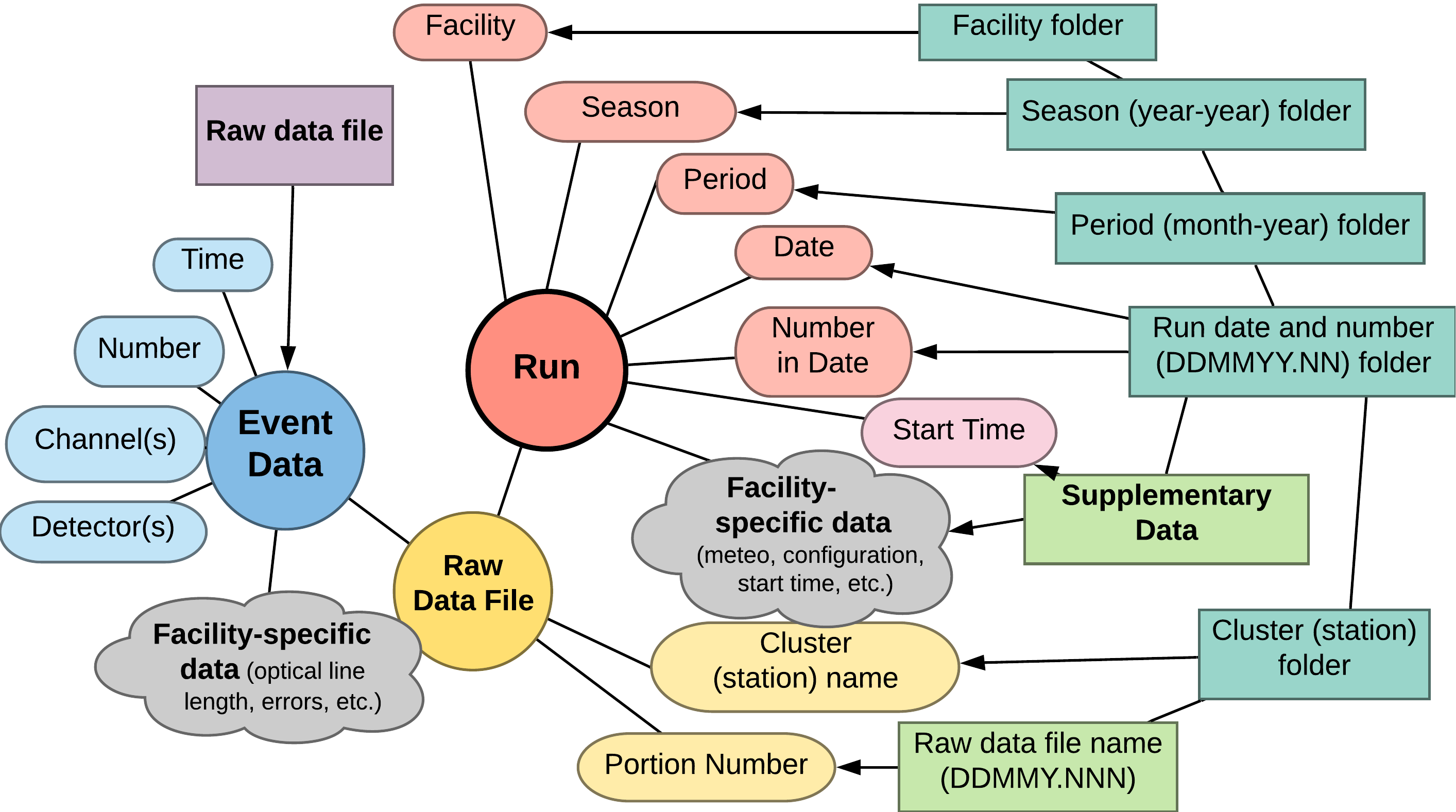}
	\label{fig:md2}
	\caption{General metadata hidden in TAIGA raw data.}
\end{figure}

Moreover, the runs and events are characterized by some environmental data.
Each run folder can be accompanied by various supplementary files with a facility-specific description of 
equipment configuration, triggering, synchronization, errors, calibration (e.g. pedestal, current, count rate), and meteorological measuring.
Some facility-specific attributes (e.g. stop-trigger position, detector number, optical line length, error package status) are also contained in raw data files.
Fig.~\ref{fig:md2} shows which of the general and facility-specific attributes can be extracted from raw data to describe events and runs.

%\begin{figure}[t]
%	\centering
%		\includegraphics[width=1.0\textwidth]{md3.pdf}
%	\label{fig:md3}
%	\caption{Facility-specific metadata hidden in TAIGA raw data.}
%\end{figure}

%\begin{figure}[t]
%	\centering
%		\includegraphics[width=0.75\textwidth]{md4.pdf}
%	\label{fig:md4}
%	\caption{Facility-specific metadata hidden in TAIGA raw data.}
%\end{figure}

%\subsection{Derived metadata}

Another part of metadata can be derived by processing TAIGA raw data to describe the following properties:
validity (determining corrected and corrupted chunks of data), reliability (calculating check-sums), availability (checking whether downloading data from the storage is possible),
accessibility (specifying user rights), popularity (registration of unique user requests and downloads). 
Moreover, some artificial neural network models~\cite{Postnikov2019} also can enrich metadata with knowledge on types of detected particles and energy.

In the general case, one event can be represented by several sequences of bytes in different files produced by one run of a facility. 
Such parts should be aggregated into events to be appropriate for the purposes of classification, analysis and use. 
The derived metadata can describe how an event is composed of parts. 
These metadata can be separated into three levels as follows: L1 --- files, L2 --- parts of events linked with files, and L3 --- events linked with parts.

The extracted and derived metadata can populate the catalog of the astroparticle data storage that we are developing. 
Thanks to these metadata, user queries to TAIGA raw data in two dimensions of the following form become possible:
\begin{verbatim}
GET data WHERE time ==
range = time between start and end (less than a night)
run = a specified run | a calibration run
night = a specified date
moonless month = a period of time (not calendar month)
summer = a summer period of time

GET data WHERE equipment ==
facility = a specified facility 
cluster = a specified cluster (station) of a facility
\end{verbatim}

\section{Metadata extractor}

\begin{figure}[t]
	\centering
		\includegraphics[width=1.0\textwidth]{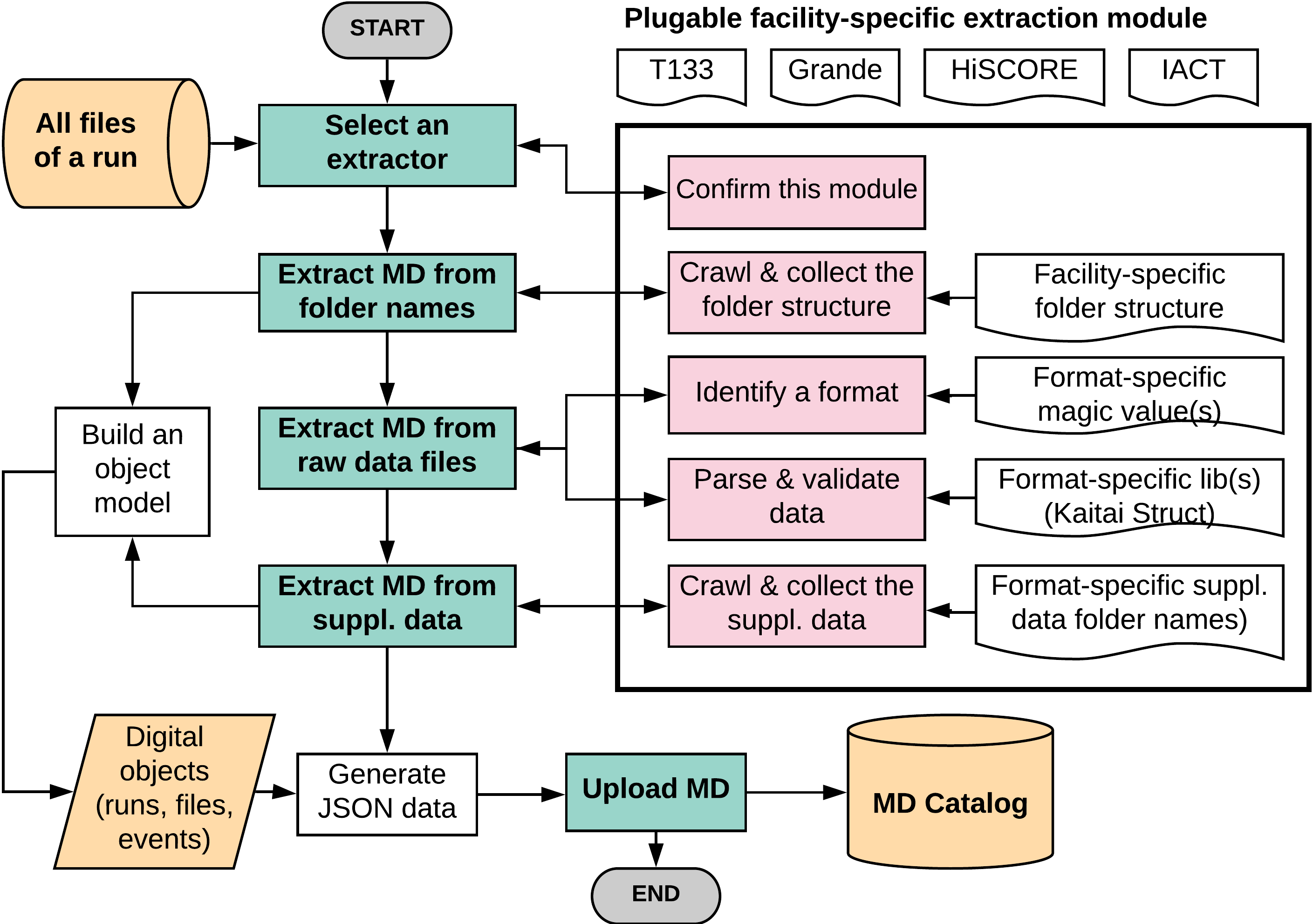}
	\label{fig:md_extractor_wf}
	\caption{Workflow for the metadata extractor.}
\end{figure}

We propose a concept of the metadata extractor for harvesting attributes of events and runs from binary files of some facility-specific formats.
It implements an extensible architecture with pluggable facility-specific modules (add-ons) as shown in Fig.~\ref{fig:md_extractor_wf}. 
Such modules can be developed based on a framework for binary data format description, e.g. ``Kaitai Struct'' or ``FlexT''. 
In this case, the considered development includes the following steps: 
exploring raw data, writing file format specifications, 
generating the source code of the software libraries for the binary file parsing,
incorporating the generated source code in the corresponding facility-specific module.

Fig.~\ref{fig:md_extractor_wf} shows the workflow for the metadata extractor. 
The workflow starts with selecting a module that is appropriate to process the input raw data in a facility-specific format.
The selected module crawls the input structure of folders and files to collect attributes being available in the folder/file names.
It identifies the format of each input file, parses and validates binary data by using an appropriate format-specific library to extract metadata from package headers. 
The module also collects attributes from the input supplementary files (e.g. facility configuration file).
All extracted metadata are used to build instances of an event object model. 
Finally, the extractor generates JSON data from these instances to upload them into the metadata catalog.

%\section{Framework}
Preliminary, we used ``Kaitai Struct'' for formally describing the file formats of TAIGA experiment~\cite{Bychkov2018b}. 
The implemented format specifications allowed us to generate source code in C/C++, Python, Java for parsing and validating binary data. 
The libraries were tested on real data: 89K files of Tunka-133, Tunka-Grande, Tunka-Rex and 120K files of TAIGA-HiSCORE, and TAIGA-IACT formats.
They can be adopted to be used in the workflow for the metadata extractor.

%\section{Metadata catalogue}
The metadata extractor can be incorporated in the micro-service architecture of the distributed storage of astroparticle data~\cite{Kryukov2018d,Kryukov2019d} we are developing.
The architecture involves placing instances of the metadata extractor on file storage nodes locally.
This allows an instance to request raw data through CernVM\footnote{\url{https://cernvm.cern.ch/portal/filesystem}} file system without their transferring among nodes of the distributed storage.
Each operating instance populates the centralized catalog with extracted metadata.
The interaction with the catalog is provided by GraphQL\footnote{\url{https://graphql.org}} API (application programming interface).
The architecture implements this interface via Graphene-Python\footnote{\url{https://graphene-python.org}} library.
It also uses the object-relational mapping based on SQLAlchemy\footnote{\url{https://www.sqlalchemy.org}} on the catalog side.
Since all digital objects (events and runs) we consider are characterized by time, 
the design of the architecture suggests to use TimeScale\footnote{\url{https://www.timescale.com}}, a time series database management system, for organizing metadata stored in the catalog.

\section{Conclusion and further work}

The best practices of scientific data maintenance recommend keeping raw data. 
This ensures facilitating reproducibility of published results and future reuse with an advanced data analysis and processing. 
The TAIGA experiment produces and accumulates a large volume of raw astroparticle data. 
To be available for the scientific community they should be accompanied by metadata with a unified interface of access. 

In our case, the important part of the metadata hidden and scattered in raw data.
Such metadata should be extracted from binary files, transformed to a unified form of digital objects, and loaded into the catalog.
To address this challenge we have developed a concept of the metadata extractor that can be extended by facility-specific extraction modules.
The extractor is aimed to automatically collect descriptive metadata from raw data files of all TAIGA formats.

Further work for the incorporation of metadata in the astroparticle data life cycle requires the following steps:
unifying the terminology (conforming a thesaurus); 
determining a set of user requests to the metadata catalog;
determining a set of hidden and derived attributes describing the digital objects;
implementing the metadata extractor;
developing the metadata catalog implementing a unified interface of access.

We believe that metadata will be useful on all stages of the astroparticle data life cycle we consider in our initiative. 
Metadata can also simplify the software development for astroparticle data exchanging and aggregation from various sources in the case of multi-messenger analysis. 
We plan to share our experience of extracting metadata from raw data with other scientific collaborations.

\section{Acknowledgments}

This work was financially supported by the Russian Scientific Foundation (Grant No 18-41-06003).

%\bibliographystyle{splncs04}
\bibliographystyle{ieeetr}
\bibliography{mybibfile}

\end{document}